\documentclass[12pt]{article}
\usepackage{color}
\usepackage{graphicx}
\usepackage{slashed}
\usepackage{amsmath,amstext,amsgen,amsbsy,amsopn,amsfonts,theorem}

\usepackage{geometry}
\begin{document}
\title{Double charmonium productions in electron-positron annihilation using Bethe-Salpeter approach}
\author{Hluf Negash$^{1}$ and Shashank Bhatnagar$^{2}$}
\maketitle
$^{1}$Department of Physics, Samara University,P.O.Box 132, Samara, Ethiopia\\
$^{2}$Department of Physics, University Institute of Sciences,
Chandigarh University, Mohali-140413, India\\
\textbf{Abstract}\\
We calculate the double charmonium production cross section within
the framework of $4\times 4$ Bethe-Salpeter Equation in the
electron-positron annihilation, at center of mass energy
$\sqrt{s}= 10.6$GeV, that proceeds through the exchange of a
single virtual photon. In this calculation, we make use of the
full Dirac structure of 4D BS wave functions of these charmonia,
with the incorporation of all the Dirac covariants (both leading
and sub-leading). The calculated cross-sections for the double
charmonium productions for final states, ($J/\Psi, \eta_c$);
($\Psi',\eta_c$) ; ($J/\Psi, \eta_c'$); and ($\Psi',\eta_c'$) are
close to experimental data, and in broad agreement with results of
other theoretical models.

\section{Introduction}
One of the challenging problems in heavy-quark physics is the
process of double charmonium production in electron-positron
annihilation at B-factories \cite{belle10,babar09,cleo01,olive14}.
Many studies  of double quarkonium production process
\cite{din,frey,ebert6,bra5,ebert9} have been preformed in order to
understand this process, and thereby the internal structures of
quarkonia and interactions of the quark and anti-quark inside the
quarkonium. In recent years the quarkonium production has been
studied in various processes at B-factories whose
measurements were made by Babar and Belle collaborations
\cite{belle10,babar09,cleo01,olive14}. Among them, the study of
charmonium production in $e^{+}e^{-}$ annihilation is particularly
interesting in testing the quarkonium production mechanisms at
center of mass energies $\sqrt{s}= 10.6GeV$. The investigation of
double charmonium production is very important since these
charmonium can be easily produced in experiments and hence their
theoretical prediction can verify the discrepancy between
different theoretical models and experimental data. As regards the
dynamical framework, to investigate the double charmonium
production is concerned, many approaches have been proposed to
deal with the cross - section of double charmonium production
\cite{din,frey,ebert6,bra5,ebert9} and the pseudoscalar and vector
charmonium production process has been recently studied in a
Bethe-Salpeter formalism \cite{xin8,elias12}. However, in these
studies the complete Dirac structure of P (pseudoscalar)and V
(vector) quarkonia was not taken into account, and calculations
were performed by taking only the leading Dirac structures,
$\gamma_5$, and $i\gamma.\varepsilon$ in the BS wave functions of
pseudoscalar and vector charmonia respectively.

\bigskip
In these calculations, we make use of the Bethe-Salpeter equation
(BSE) approach
\cite{mitra01,bhatnagar92,mitra99,bhatnagar91,bhatnagar14}, which
is a conventional approach in dealing with relativistic bound
state problems. Due to its firm base in quantum field theory, and
being a dynamical equation based approach, it provides a realistic
description for analyzing hadrons as composite objects, and can be
applied to study not only the low energy hadronic processes, but
also the high energy production processes involving quarkonia as
well.

\bigskip
In our recent works \cite{hluf15,hluf16,hluf17,hluf}, we employed
BSE under Covariant Instantaneous Ansatz, which is a Lorentz-
invariant generalization of Instantaneous Approximation, to
investigate the mass spectra and the transition amplitudes for
various processes involving charmonium and bottomonium. The BSE
framework using phenomenological potentials can give consistent
theoretical predictions as more and more data are being
accumulated. In our studies on $4\times 4$ BSE, in all processes
except in \cite{hluf17}, the quark - anti quark loop involved a
single hadron-quark vertex, which was simple to handle. However
for the transitions such as $V\rightarrow P+\gamma$ (where $V$
vector and $P$ pseudoscalar quarkonium), which we have studied in
\cite{hluf17} and for the process of double charmonium production,
$e^{+}e^{-}\rightarrow V + P$, we will study in present work, the
process requires calculation involving two hadron-quark vertices,
due to which the calculation becomes more and more difficult to
handle. However in the present work, we demonstrate an explicit
mathematical procedure for handling this problem using the
formulation of $4\times 4$ Bethe - Salpeter Equation under
Covariant Instantaneous Ansatz. We will use this framework for the
calculation of cross section for the production of double
charmonia, $(J/\psi, \eta_{c}), (J/\psi, \eta'_{c}),
(\psi',\eta_{c})$, and $(\psi',\eta'_{c})$ in electron-positron
annihilation that proceeds through a single virtual photon at
center of mass energies, $\sqrt{s}= 10.6$ GeV., where such
problems do not enter in our previous papers on $4\times 4$ BSE.
\cite{hluf15,hluf16,hluf17,hluf,hluf19,bhatnagar18}.

\bigskip
This paper is organized as follows. In section \textbf{2}, We
introduce the detailed formulation of the transition amplitudes for
the process of double charmonium production and the numerical
results of cross - sections. Finally, we give the discussions and
conclusions in section \textbf{3}.

\section{Formulation of double charmonium production amplitude}
We start from the lowest order Feynman diagrams for the process,
$e^{+}e^{-}\rightarrow V + P$, as given in Fig.1. There are four
Feynman diagrams for the production of double charmonium, one of
which is shown in Fig.1, while the other three can be obtained by
reversing the arrows of the quark lines.

\begin{figure}[h]
\centering
  \includegraphics[width=12cm,]{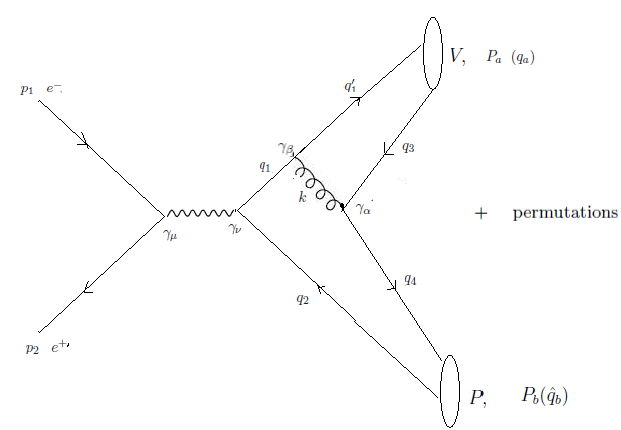}
    \caption{Leading order Feynman diagrams for the production of double charmonium in
     $e^{+}e^{-}$- annihilation.}\label{4}
\end{figure}

 The relativistic amplitude $M^{1}_{fi}$ for double charmonium production, corresponding to Fig.1, is given by the
one-loop momentum integral as:
\begin{equation}
M^{1}_{fi}= \frac{2^{7}\pi^{2}\alpha_{em}\alpha_{s}}{3^{2}s}
\left[\overline{\nu}(p_{2})\gamma_{\mu}u(p_{1})\right]
\int\frac{d^{4}q_{a}}{(2\pi)^{4}}\int\frac{d^{4}q_{b}}{(2\pi)^{4}}
Tr[\overline{\Psi}(P_{a},q_{a})\gamma_{\beta}S_{F}(q_{1})\gamma_{\mu}\overline{\Psi}(P_{b},q_{b})\gamma_{\beta}]
\frac{1}{k^{2}}
\end{equation}
where, $s$ is the Mendelstam variable defined as,
$s=-(p_{1}+p_{2})^{2}$, $\alpha_{em}=\frac{e^{2}}{4\pi}$ is called
the electromagnetic coupling constant and
$\alpha_{s}=\frac{g^{2}_{s}}{4\pi}$ is the strong coupling strength,
$\overline{\Psi}(P_{a},q_{a})$ and $\overline{\Psi}(P_{b},q_{b})$
are the conjugations of the BS wave function of vector and
pseudoscalar charmonium respectively. From the figure, we can relate
the momenta of the quark and anti-quark respectively as:
\begin{eqnarray}
&&\nonumber q'_{1}=\frac{1}{2}P_{a}+q_{a}\\&& \nonumber
q_{3}=\frac{1}{2}P_{a}-q_{a}\\&&
 \nonumber q_{4}=\frac{1}{2}P_{b}+q_{b}\\&&
 q_{2}=\frac{1}{2}P_{b}-q_{b}
\end{eqnarray}
and the momenta for the propagators of gluon and quark are given
respectively by:
\begin{eqnarray}
&&\nonumber k=\frac{1}{2}(P_{a}+P_{b})-q_{a}+q_{b}\\&&
q_{1}=P_{a}+\frac{1}{2}P_{b}+q_{b}
\end{eqnarray}
As the quark and gluon propagators depend upon the internal hadron
momenta $q_{a}$ and $q_{b}$, the calculation of amplitude is going
to involve integrations over these internal momenta and will be
quite complex. Hence following Ref. \cite{hluf15,hluf16,hluf17},
we simplify the calculation, by reducing the 4-dimensional
expression of BS amplitude, $M^{1}_{fi}$ into 3-dimensional
expression of BS amplitude, $M^{1}_{fi}$ and employing the heavy
quark approximation on the propagators, the momenta $k$ and
$q_{1}$ can be written as, $k\approx\frac{1}{2}(P_{a}+P_{b})$ and
$q_{1}\approx P_{a}+\frac{1}{2}P_{b}$, which leads to,
$k^{2}\approx\frac{s}{4}$ and
$q^{2}_{1}\approx\frac{s}{2}+m_{c}^{2}$ \cite{xin8,elias12}. Then,
with the above approximation and applying the definition of 3D BS
wave function in \cite{hluf15,hluf16,hluf17,hluf}
$\psi(\hat{q}_{a}) = \frac{i}{2\pi }\int
M_{a}d\sigma\Psi(P_{a},q_{a})$ and $\psi(\hat{q}_{b}) =
\frac{i}{2\pi }\int M_{b}d\sigma_{b}\Psi(P_{b},q_{b})$, the double
charmonium production BS amplitude, $M^{1}_{fi}$ can be written in
the instantaneous Bethe-Salpeter amplitude form as:
\begin{equation}
M^{1}_{fi}=-\frac{2^{10}\pi^{2}\alpha_{em}\alpha_{s}}{3^{2}s^{3}}
\left[\overline{\nu}(p_{2})\gamma_{\mu}u(p_{1})\right]
\int\frac{d^{3}\widehat{q}_{a}}{(2\pi)^{3}}\int\frac{d^{3}\widehat{q}_{b}}{(2\pi)^{3}}
Tr[\overline{\psi}(\widehat{q}_{a})\gamma_{\beta}(-i\slashed{P_{a}}-\frac{i}{2}\slashed{P_{b}}+m_{c})
\gamma_{\mu}\overline{\psi}(\widehat{q}_{b})\gamma_{\beta}]
\end{equation}
where $\psi(\hat{q}_{a})$ and $\psi(\hat{q}_{b})$ are the
relativistic BS wave function of pseudoscalar and vector charmonium
respectively. We now give details of calculation of double
charmonium production for the process, $e^{+}e^{-}\rightarrow V + P$
in the next section.

\begin{itemize}
\item \textbf{For the production process, $e^{+}+e^{-}\rightarrow V + P$}\\
The relativistic BS wave function of pseudoscalar and vector
charmonium are taken from our recent papers
\cite{hluf15,hluf16,hluf17} respectively as:
\begin{equation}
\psi_{P}(\hat{q}_{b}) =N_{P}
[M_{b}+\slashed{P}_{b}+\frac{\slashed{\hat{q}_{b}}\slashed{P}_{b}}{m_{c}}]\gamma_{5}\phi_{P}(\hat{q}_{b})
\end{equation}
for $P_{b}$ and $M_{b}$, is the momentum and mass of the
pseudoscalar charmonium respectively and $N_{P}$ is the BS
normalization of the pseudoscalar charmonium, which is given in a
simple form as:
\begin{equation}
N_{P}=\left[\frac{16M_{b}}{m_{c}}\int\frac{d^{3}\hat{q}_{b}}{(2\pi)^{3}}
\frac{\hat{q}_{b}^{2}}{\omega_{b}}\phi_{P}^{2}(\hat{q}_{b})\right]^{-1/2}
\end{equation}
and
\begin{equation}
\psi_{V}(\hat{q}_{a})=N_{V}[M_{a}\slashed{\varepsilon}+\hat{q}_{a}.\varepsilon\frac{M_{a}}{m_{c}}
+\slashed{\varepsilon}\slashed{P}_{a}+
\frac{\slashed{P}_{a}\hat{q}_{a}.\varepsilon}{m_{c}}-\frac{\slashed{P}_{a}\slashed{\varepsilon
}\slashed{\hat{q}_{a}}}{m_{c}}]\phi_{V}(\hat{q}_{a})
\end{equation}
were, $\varepsilon$ is the polarization vector of the vector
quarkonia, $P_{a}$ and $M_{a}$, is the momentum and mass of the
vector quarkonia respectively and $N_{V}$ is the BS normalization of
the vector quarkonia, which is given in a simple form as:
\begin{equation}
 N_{V} =\left[16m_{c}M_{a}\int\frac{d^{3}\hat{q}_{a}}{(2\pi)^{3}}
\frac{\hat{q}_{a}^{2}}{\omega_{a}^{3}}\phi_{V}^{2}(\hat{q}_{a})\right]^{-1/2}
\end{equation}
where $\phi_{P}(\hat{q}_{b})$ and $\phi_{V}(\hat{q}_{a})$ are the
radial wave functions of $P$ and $V$, which are solutions of the
3D BSE for pseudoscalar and vector quarkonia respectively (see
Ref.\cite{hluf16}). The adjoint BS wave functions for pseudoscalar
and vector charmonia are obtained from
$\overline{\psi}_{P,V}(\hat{q}_{b,a})=\gamma^{0}(\psi_{P,V}(\hat{q}_{b,a}))^{+}\gamma^{0}$.
With the substitution of adjoint BS wave functions of $P$ and $V$
charmoni into the 3D amplitude Eq.(4), $M^{1}_{fi}$ becomes:
\begin{equation}
M^{1}_{fi}=-\frac{2^{10}\pi^{2}\alpha_{em}\alpha_{s}}{3^{2}s^{3}}
\left[\overline{\nu}(p_{2})\gamma_{\mu}u(p_{1})\right]
\int\frac{d^{3}\widehat{q}_{a}}{(2\pi)^{3}}N_{V}\phi_{V}(\hat{q}_{a})\int\frac{d^{3}\widehat{q}_{b}}{(2\pi)^{3}}N_{P}\phi_{P}(\hat{q}_{b})
[TR]
\end{equation}

were

\begin{equation}
[TR]=Tr[(-M_{a}\slashed{\varepsilon}+\hat{q}_{a}.\varepsilon\frac{M_{a}}{m_{c}}-
\slashed{\varepsilon}\slashed{P}_{a}-
\frac{\slashed{P}_{a}\hat{q}_{a}.\varepsilon}{m_{c}}+\frac{\slashed{\hat{q}_{a}}\slashed{\varepsilon
}\slashed{P}_{a}}{m_{c}})\gamma_{\beta}(-i\slashed{P_{a}}-\frac{i}{2}\slashed{P_{b}}+m_{c})
\gamma_{\mu}(M_{b}+\slashed{P}_{b}+\frac{\slashed{\hat{q}_{b}}\slashed{P}_{b}}{m_{c}})\gamma_{5}\gamma_{\beta}].
\end{equation}
Applying the trace theorem and evaluating trace over the gamma
matrices, one can obtain the expression:
\begin{eqnarray}
\left[TR\right]= &&\nonumber
-8M_{a}\epsilon_{\nu\delta\mu\lambda}\varepsilon_{\nu}P_{a\lambda}P_{b\delta}
+\frac{8i}{m_{c}}\left[(\hat{q}_{a}.\varepsilon)\epsilon_{\lambda\delta\sigma\mu}P_{a\lambda}P_{b\beta}\hat{q}_{b\sigma}
+\epsilon_{\rho\nu\lambda\phi}(P_{b\phi}\hat{q}_{b\mu}-P_{b\mu}\hat{q}_{b\phi})\right]\\&&
\nonumber\frac{8(\hat{q}_{a}.\varepsilon)
M_{a}}{m_{c}^{2}}\left[\epsilon_{\mu\sigma\beta\delta}P_{a\beta}P_{b\delta}\hat{q}_{b\sigma}+
\epsilon_{\lambda\beta\mu\sigma}P_{a\lambda}P_{b\beta}\hat{q}_{b\sigma}\right]
+\frac{16M_{b}^{2}}{m_{c}}
\epsilon_{\nu\delta\sigma\mu}\varepsilon_{\nu}P_{b\delta}\hat{q}_{b\sigma}\\&&
-\frac{8(\hat{q}_{a}.\varepsilon)}{m_{c}}\epsilon_{\lambda\delta\sigma\mu}P_{a\lambda}P_{b\delta}\hat{q}_{b\sigma}
+8iM_{b}\epsilon_{\rho\nu\lambda\mu}\varepsilon_{\nu}P_{a\lambda}\hat{q}_{a\rho}
\end{eqnarray}
After some mathematical steps, we can get the following
expression:
\begin{equation}
\left[TR\right]=-16m_{c}\epsilon_{\mu\nu\rho\sigma}\varepsilon_{\mu}P_{a}^{\rho}P_{b}^{\sigma}
\end{equation}
Thus we can express the amplitude $M^{1}_{fi}$ for the vector and
pseudoscalar charmonium production as:
\begin{equation}
M^{1}_{fi}=\frac{2^{14}\pi^{2}\alpha_{em}\alpha_{s}m_{c}}{3^{2}s^{3}}
\left[\overline{\nu}(p_{2})\gamma_{\mu}u(p_{1})\right]\epsilon_{\mu\nu\rho\sigma}\varepsilon_{\mu}P_{a}^{\rho}P_{b}^{\sigma}
\int\frac{d^{3}\widehat{q}_{a}}{(2\pi)^{3}}N_{V}\phi_{V}(\hat{q}_{a})
\int\frac{d^{3}\widehat{q}_{b}}{(2\pi)^{3}}N_{P}\phi_{P}(\hat{q}_{b})
\end{equation}
The full amplitude for the process $e^{+}+e^{-}\rightarrow V+P$, can
be obtained by summing over the amplitudes of all the possibilities
in Fig. 1. Then, the unpolarized total cross-section is obtained by
summing over various $V+P$, spin-states and averaging over those of
the initial state $e^{+}e^{-}$, which is given as:
\begin{equation}
\sigma=\frac{1}{32\pi}\frac{\sqrt{s-16m^{2}_{c}}}{s^{3/2}}\int\frac{1}{4}
\sum_{spin}|M_{total}|^{2}d\cos\theta
 \end{equation}
The total amplitude, $|M_{total}|^{2}$ is given as:
\begin{eqnarray}
\frac{1}{4}
\sum_{spin}|M_{total}|^{2}=\frac{2^{30}\pi^{4}\alpha_{em}^{2}\alpha_{s}^{2}m^{2}_{c}}{3^{4}s^{5}}
\left[\int\frac{d^{3}\widehat{q}_{a}}{(2\pi)^{3}}N_{V}\phi_{V}(\hat{q}_{a})\right]^{2}
\left[\int\frac{d^{3}\widehat{q}_{b}}{(2\pi)^{3}}N_{P}\phi_{P}(\hat{q}_{b})\right]^{2}
\end{eqnarray}
After integration over the phase space, the total cross-section
for vector and pseudoscalar charmonium production is:
\begin{equation}
  \sigma_{e^{+}e^{-}\rightarrow VP}=\frac{2^{27}\pi^{3}\alpha_{em}^{2}\alpha_{s}^{2}m^{2}_{c}}
  {3^{4}s^{4}}\left(1-\frac{16m^{2}_{c}}{s}\right)^{\frac{3}{2}}
  \left[\int\frac{d^{3}\widehat{q}_{a}}{(2\pi)^{3}}N_{V}\phi_{V}(\hat{q}_{a})\right]^{2}
\left[\int\frac{d^{3}\widehat{q}_{b}}{(2\pi)^{3}}N_{P}\phi_{P}(\hat{q}_{b})\right]^{2}
\end{equation}
The algebraic expressions of the wave functions of pseudoscalar
and vector charmonium, $\phi_{P(V)}(\hat{q}_{b(a)})$, for ground
(1S) and first excited (2S) states respectively, that are obtained
as analytic solutions of the corresponding mass spectral equations
of these quarkonia in an approximate harmonic oscillator basis are
\cite{hluf16}:
\begin{eqnarray}
&&\nonumber
\phi_{P(V)}(1S,\hat{q})=\frac{1}{\pi^{3/4}\beta_{P(V)}^{3/2}}e^{-\frac{\hat{q}^{2}}{2\beta_{P(V)}^{2}}}\\&&
\phi_{P(V)}(2S,\hat{q})=
\frac{\sqrt{3/2}}{3\pi^{3/4}\beta_{P(V)}^{7/2}}\left(3\beta_{P(V)}^{2}-2\hat{q}^{2}\right
)e^{-\frac{\hat{q}^{2}}{2\beta_{P(V)}^{2}}},
\end{eqnarray}
where the inverse range parameter $\beta_{P}$ for pseudoscalar
charmonium is
$\beta_{P}=(4\frac{m_{c}\omega^{2}_{q\bar{q}}}{\sqrt{1+2A_{0}(N+\frac{3}{2})}})^{\frac{1}{4}}$,
while $\beta_{V}$ for vector charmonium is
$\beta_{V}=(2\frac{m_{c}\omega^{2}_{q\bar{q}}}{\sqrt{1+2A_{0}(N+\frac{3}{2})}})^{\frac{1}{4}}$,
and these two constants depend on the input parameters and contain
the dynamical information, and they differ from each other due to
spin-spin interactions. It can be checked that our cross sectional
formula in Eq.(16) scales as $\frac{\alpha_{em}^2\alpha_s^2
m^6}{s^4}$, with $m$ being the mass of $c$-quark, where in
Eq.(16), the wave functions, $\phi_{(P,V)}$ involve the inverse
range parameters, $\beta_{P,V} \sim  m^{1/2}$.

\end{itemize}
\textbf{Numerical results}\\
We had calculated recently the mass spectrum and various decays of
ground and excited states of pseudoscalar and vector quarkonia in
\cite{hluf16,hluf17,hluf}. The same input parameters are employed in
this calculation as in our recent works,  given in table 1.
\begin{table}[h]
  \begin{center}
  \renewcommand{\tabcolsep}{12pt}
\begin{tabular}{ccccc}
\hline
     $C_{0}$ &$\omega_{0}(GeV)$&$\Lambda(GeV)$&$A_{0}$&$m_{c}(GeV)$\\ \hline
    0.210&0.150&0.200&0.010&1.490\\
    \hline
      \end{tabular}
      \end{center}
  \caption{Input parameters for this study}
    \end{table}\\
Using these input parameters listed in Table 1, we calculate the
cross sections of pseudoscalar and vector charmonium production in
our framework are listed in Table 2.
\begin{table}[htbp]
   \begin{center}
   \renewcommand{\tabcolsep}{8pt}
\begin{tabular}{llllllll}
  \hline
  Production Process&$\sigma$(Our result)&$\sigma$\cite{belle10}&$\sigma$\cite{babar09}&$\sigma$\cite{bra5}
  &$\sigma$\cite{ebert9}  &$\sigma$\cite{xin8}&$\sigma$\cite{elias12} \\\hline
    $e^{+}e^{-}\rightarrow J/\psi\eta_{c}$&20.770&25.6$\pm$2.8&
    17.6$\pm$2.8&26.7&22.2$\pm$4.2&22.3&21.75 \\
     $e^{+}e^{-}\rightarrow \psi'\eta_{c}$&11.643&16.3$\pm$4.6
    &&16.3&15.3$\pm$2.9&& \\
    $e^{+}e^{-}\rightarrow J/\psi\eta'_{c}$&11.177&16.5$\pm$3.0&
    16.4$\pm$3.7&26.6 &16.4$\pm$3.1&&\\
    $e^{+}e^{-}\rightarrow \psi'\eta'_{c}$&6.266&16.0$\pm$5.1
    &&14.5&9.6$\pm$1.8&& \\
     \hline
     \end{tabular}
   \end{center}
   \caption{The total cross sections of pseudoscalar and vector charmonium production for ground
    and first excited states
    with experimental data and other theoretical models(in units of fb).}
 \end{table}

We wish to mention here that the $b\overline{b}b\overline{b}$ as
well as $b\overline{b}c\overline{c}$ production has not been
observed so far, but cross sections have been predicted for $e^-
e^+\rightarrow \Upsilon +\eta_b$ at center of mass energy,
$\sqrt{s}= 25- 30$ GeV. We thus wished to check our calculations
for double bottomonium ($\Upsilon\eta_{b}$) production in
electron-positron annihilation for sake of completeness. The same
can be studied with our framework used in this work with little
modifications. Taking the mass of the $b-$quark that is fixed
 from our recent work on spectroscopy of $b\overline{b}$ states as $m_{b}=5.07$GeV
 \cite{hluf16}, and other input parameters the same, the cross
 section for production of ground states of $\eta_b$, and
 $\Upsilon$ for energy range ($\sqrt{s}=28-30$)GeV is given in Table
 3.
 \begin{table}[htbp]
   \begin{center}
   \renewcommand{\tabcolsep}{12pt}
\begin{tabular}{lllll}
  \hline
           Production Process&$\sigma$(Our result)&$\sigma$\cite{belle10}&$\sigma$\cite{babar09}&$\sigma$\cite{xin8}for ($\sqrt{s}=25-30$)GeV\\\hline
     $\sigma[e^{+}e^{-}\rightarrow\Upsilon\eta_{b}]$&0.155-0.058&               &                    &0.16-0.06\\
     \hline
     \end{tabular}
   \end{center}
   \caption{The total cross sections of pseudoscalar and vector bottomonium production for their ground
    states with predictions of other theoretical models (in units of fb).}
\end{table}

 \section{Discussions and conclusion}
Our main aim in this paper was to study the cross section for
double charmonium production in electron-positron collisions at
center of mass energies, $\sqrt{s}= 10.6 GeV.$, having
successfully studied the mass spectrum and a range of low energy
processes using an analytic treatment of $4 \times 4$
Bethe-Salpeter Equation
(BSE)\cite{hluf16,hluf17,hluf,hluf19,bhatnagar18}. This is due to
the fact any quark model model should successfully describe a
range of processes- not only the mass spectra, and low energy
hadronic decay constants/decay widths, but also the high energy
production processes involving these hadrons, and all within a
common dynamical framework, and with a single set of input
parameters, that are calibrated to the mass spectrum.

Further, the exclusive production of double heavy charmonia in
$e^+ e^-$ annihilation has been a challenge to understand in
heavy-quark physics, and has received considerable attention in
recent years, due to the fact that there is a significant
discrepancy in the data of Babar\cite{babar09} and
Belle\cite{belle10}, and the calculations performed in
NRQCD\cite{din,frey}, of this process at $\sqrt{s}=10.6 GeV.$
Using Leading order (LO) QCD diagrams alone, lead to cross
sections that are an order of magnitude smaller than data
\cite{babar09,belle10,olive14}. These discrepancies were then
resolved by taking into account the Next-to-Leading Order (NLO)
QCD corrections combined with relativistic corrections
\cite{braaten03,zhang08}), though the NLO contributions were
larger than the LO contributions \cite{xin8}.

This process has also been studied in Relativistic Quark Model
\cite{ebert6,ebert9}, Light Cone formalism \cite{bra5}, and
Bethe-Salpeter Equation \cite{xin8,elias12}. Relativistic
corrections to cross section in Light-Cone formalism (in
\cite{bra5}) were considered to eliminate the discrepancy between
theory and experiment. Further in an attempt to explain a part of
this discrepancy, \cite{bodwin03} even suggested that processes
proceeding through two virtual photons may be important. In
Relativistic quark model \cite{ebert6}, which incorporated
relativistic treatment of internal motion of quarks, and bound
states, improvements in the results were obtained.

We wish to mention that in our framework of $4\times 4$
Bethe-Salpeter Equation (BSE) using the Covariant Instantaneous
Ansatz, where we treat the internal motion of quarks and the bound
states in a relativistically covariant manner, we obtained results
on cross sections (in Table 2) for production of opposite charge
parity $c\overline{c}$ states, such as, $J/\Psi, \eta_c$;
$\Psi',\eta_c$ ; $J/\Psi, \eta_c'$; and $\Psi',\eta_c'$, that are
close to central values of data \cite{babar09, belle10, olive14}
using leading order (LO) QCD diagrams alone. This validates the
fact that relativistic quark models such as BSE are strong
candidates for treating not only the low energy processes, but
also the high energy production processes involving double heavy
quarkonia, due to their consistent relativistic treatment of
internal motion of quarks in the hadrons, where our BS wave
functions that take into account all the Dirac structures in
pseudoscalar and vector mesons in a mathematically consistent
manner, play a vital role in the dynamics of the process. We also
give our predictions on cross section for double $b\overline{b}$
production at energies 28- 30 GeV. in Table 3 for future
experiments at colliders.

The main objective of this study was to test the validation of our
approach, which provides a much deeper insight than the purely
numerical calculations in $4\times 4$ BSE, that are very common in
the literature. We wish to mention that, we have not encountered
any work in $4\times 4$ representation of BSE, that treats this
problem analytically. On the contrary all the other $4\times 4$
BSE approaches adopt a purely numerical approach of solving the
BSE. We are also not aware of any other BSE framework, involving
$4\times4$ BS amplitude, and with all the Dirac structures
incorporated in the 4D hadronic BS wave functions (in fact many
works used only the leading Dirac structures, for instance see
Ref.\cite{xin8,elias12}) for calculations of this production
process. We treat this problem analytically by making use of the
algebraic forms of wave functions, $\phi_{(P,V)}$ derived
analytically from mass spectral equations \cite{hluf16}, for
calculation of cross section of this double charmonium production
process.

This calculation involving production of double charmonia in
electron-positron annihilation can be easily be extended to
studies on other processes (involving the exchange of a single
virtual photon) observed at B-factories such as, $e^- e^+ ->
\Psi(2S) \gamma$, $e^- e^+ ->J/\Psi \chi_{c0}$, and $e^- e^+
->\Psi(2S) \chi_{c0}$. We further wish to extend this study to
processes involving two virtual photons, such as the production of
double charmonia in final states with $C= +1$ such as, $e^- e^+
\rightarrow J/\Psi J/\Psi$), recently observed at BABAR.

These processes that we intend to study next, are quite involved,
and the dynamical equation based approaches such as BSE is a
promising approach. And since these processes involve
quark-triangle diagrams, we make use of the techniques we used for
handling such diagrams recently done in
\cite{hluf17,elias12a,elias13}. Such analytic approaches not only
lead to better insight into the mass spectra, and low energy decay
processes involving charmonia, but also their high energy
production processes such as the one studied here.

Note added: At proof correction stage, we come to know of about the recent experimental observation
by CMS collaboration \cite{cms} of two excited $B_c^{+}(2S)$, and $B_c^{*+}(2S)$ states in $p p$
collisions at $\sqrt{s}=13$ TeV. at Large Hadron Collider (LHC). The theoretical study of this process will be a challenge to hadronic physics.

\textbf{Acknowledgements}:\\
 This work was carried out at Samara University, Ethiopia, and at Chandigarh
 University, India. We thank these Institutions for the facilities provided during the course of
 this work.

\end{document}